\documentclass[cits]{PoS}
\def\ngc {NGC~5506}
\usepackage{epsfig}

\title{High-energy emission from NGC~5506, the brightest hard X-ray Narrow Line Seyfert 1 galaxy}

\ShortTitle{High-energy emission from NGC~5506, the brightest hard X-ray NLS1}

\author{\speaker{Simona Soldi}\\
        Laboratoire AIM - CNRS - CEA/DSM - Universit\'e Paris Diderot (UMR 7158), CEA Saclay, DSM/IRFU/SAp, F91191 Gif-sur-Yvette, France\\
        E-mail: \email{simona.soldi@cea.fr}
	}
\author{Volker Beckmann\\
	Fran\c{c}ois Arago Centre, APC, Universit\'e Paris Diderot, CNRS/IN2P3, CEA/DSM, Observatoire de Paris, 13 rue Watt, 75205 Paris Cedex 13, France
	}
\author{Neil Gehrels\\
	NASA Goddard Space Flight Center, Astrophysics Science Division, Code 661, Greenbelt, MD 20771, USA
	}
\author{Sandra De Jong\\
	Fran\c{c}ois Arago Centre, APC, Universit\'e Paris Diderot, CNRS/IN2P3, CEA/DSM, Observatoire de Paris, 13 rue Watt, 75205 Paris Cedex 13, France
 	}
\author{Piotr Lubi\'nski\\	        
        Nicolaus Copernicus Astronomical Center, Polish Academy of Sciences ul. Rabianska 8, 87--100 Torun, Poland 
	}

\abstract{We present results on the hard X-ray emission of \ngc, the brightest narrow line Seyfert 1 galaxy above 20~keV.
	  All the recent observations by \textit{INTEGRAL}, \textit{Swift} and \textit{Suzaku} have been analysed and spectral analysis
	  during nine separated time periods has been performed. While flux variations by a factor of 2 were detected during the last 7 years,
	  only moderate spectral variations have been observed, with the hint of a hardening of the X-ray spectrum and a decrease of the 
	  intrinsic absorption with time. Using \textit{Suzaku} observations it is possible to constrain the amount of Compton reflection 
	  to $R = 0.6-1.0$, in agreement with previous results on the source. The signature of Comptonisation processes can also 
	  be found in the detection of a high-energy cut-off during part of the observations, at energies $E_{\rm C} \geq 40-100 \, \rm keV$. 
	  When a Comptonisation model is applied to the \textit{Suzaku} data, the temperature and the optical depth of the Comptonising 
	  electron plasma are measured at $kT_{\rm e} = 60-80 \rm \, keV$ and $\tau = 0.6-1.0$, respectively.
	  The properties inferred for \ngc\ in this study agree with those based on other data sets for the same AGN, and fit the 
	  picture of NLS1 having in general 
	  lower high-energy cut-offs at hard X-rays than their broad line equivalent.
	  }

\FullConference{Narrow-Line Seyfert 1 Galaxies and their place in the Universe - NLS1,\\
		April 04-06, 2011\\
		Milan Italy}

\begin{document}

\section{Introduction}

In the X-rays below 10~keV, narrow line Seyfert 1 galaxies (NLS1) have characteristics that distinguish them from the broad line objects: they present 
a soft-excess, strong X-ray variability, a steep X-ray continuum ($\Gamma = 2.1-2.2$) and a sharp decrease at about 7 keV, interpreted as due to partial covering
or to reflection and light bending effects \cite{gallo06}.
On the other hand, the hard X-ray properties of NLS1 are not yet well defined, as only a handful of objects has been studied in detail up to now.
A work on a small sample of NLS1 observed with \textit{INTEGRAL} suggested the presence of a relatively low temperature of the electron plasma 
responsible for the Comptonisation at the origin of the X-ray emission \cite{malizia08}. This lower temperature (when compared to broad line Seyfert 1) is 
an indication of a more efficient cooling of the emitting plasma that could be related to the higher accretion rates of NLS1 and, therefore, to a higher density
of their accretion flow. This would fit the scenario of NLS1 representing a class of AGN that are in rapid evolution and have not yet accreted enough mass 
to have $10^{8-9} \, \rm M_{\odot}$ black holes \cite{mathur00,peterson00}.

The nearby AGN \ngc\  ($z = 0.0062$) is one of the most luminous and brightest Seyfert galaxies in the hard 
X-rays and turned out to be the brightest NLS1 in this energy band. 
Due to its brightness, it has been observed by several satellites
since the beginning of X-ray astronomy.
Even though the flux below 10~keV changed by a factor of 3.5 over the time of \textit{BeppoSAX}
observations, only little or no variations of the hard X-ray flux nor of the X-ray spectral shape were
detected \cite{dadina07}. Similarly, a study of the variability of \ngc\ with \textit{RXTE} data
indicates little energy dependence of the variability in the X-ray band below 15~keV \cite{uttley05} and a variability
rather on time scales of several months in the hard X-ray domain \cite{beckmann07}.

We present here preliminary results on the emission of \ngc\ above 20~keV as measured with
the present hard X-ray satellites, \textit{INTEGRAL} \cite{winkler03}, \textit{Swift} \cite{gehrels04} and \textit{Suzaku} \cite{mitsuda07},
which allow us to follow the behaviour of this source across 7 years of observations. 

\section{Recent hard X-ray observations}
Since 2002, \textit{INTEGRAL} has collected a total of 515~ks and 135~ks of effective exposure time on \ngc\ with its high-energy instruments
IBIS/ISGRI \cite{lebrun03} and JEM-X \cite{lund03}, respectively. All these data have been analysed using version 9.0 of the Offline Scientific Analysis 
Software (OSA) and the spectra have been extracted using the standard OSA spectral extraction.

\begin{figure}[!t] 
\hspace{0.4cm}
\includegraphics[width=.93\textwidth,angle=0]{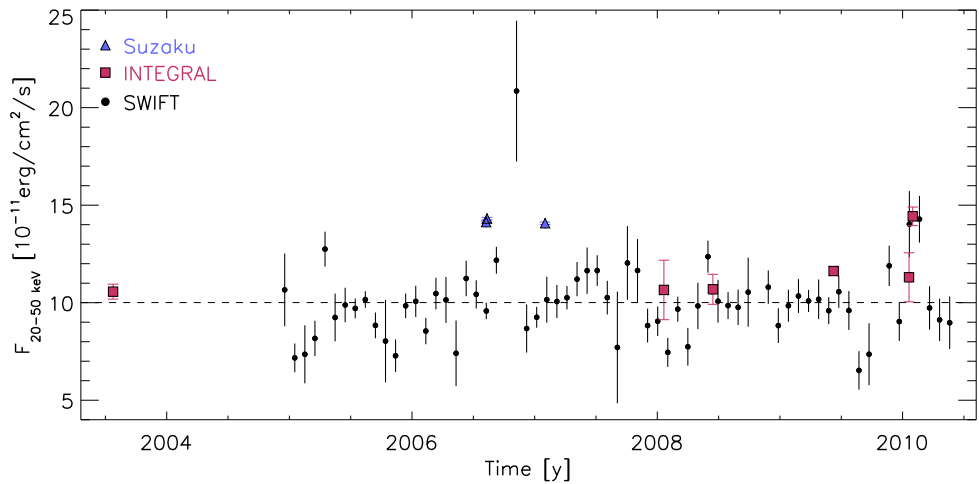} 
\caption{20--50~keV light curve of \ngc\ as observed with \textit{Swift}/BAT (binned to 30 days; black circles), 
         \textit{INTEGRAL}/IBIS/ISGRI (red squares) and \textit{Suzaku}/PIN (blue triangles).
	 The dashed line indicates the average value of the hard X-ray flux measured by BAT.
	 } 
\label{lc_ngc} 
\end{figure}

Thanks to its large field of view and to the observing strategy focused on GRB follow-ups, \textit{Swift}/BAT has monitored \ngc\ regularly for more than 5 years. 
The hard X-ray light curve of this source reported in Fig.~\ref{lc_ngc} is provided by the BAT 58-month 
survey\footnote{\emph{http://swift.gsfc.nasa.gov/docs/swift/results/bs58mon/}} \cite{baumgartner11}.
Four \textit{Swift} pointed observations of the source have been performed in May--June 2009 and January 2010, for a total of 13~ks
of dead-time corrected XRT exposure. These data have been analysed with the \textit{Swift} software version 3.4 distributed with the HEAsoft 6.7.0 
package and the latest available calibration files.

\textit{Suzaku} observed \ngc\ during three pointed observations in August 2006 and January 2007, for a total of about 150~ks effective exposure.
The XIS spectra have been extracted using the HEAsoft 6.8.0 package and the calibration files released in March 2010. 
3$\times$3 and 5$\times$5 modes events were reprocessed with the recent software and merged before the spectral extraction. 
The spectra of the front-illuminated CCDs (XIS0, XIS2 and XIS3) have been added together.
The HXD/PIN data have been reprocessed with standard cuts using the HEAsoft 6.7.0 package.

\section{Spectral analysis} 
IBIS/ISGRI observations close in time (within maximum 3 weeks) have been grouped to obtain a sufficient S/N for the spectral analysis.
Only periods where a quasi-simultaneous coverage in the soft ($<10 \rm \, keV$ with JEM-X, XRT or XIS) and hard ($> 20 \rm \, keV$
with ISGRI or PIN) X-ray band was available have been considered for the spectral analysis (Table~\ref{tab}).
The spectra have been fitted with a simple absorbed power law model, modified by a high-energy cut-off and a Compton reflection 
component ({\sc pexrav} model), when statistically required.
The disc inclination angle has been fixed to $\theta = 40^{\circ}$ \cite{guainazzi10} and the iron abundances to solar values.

The high S/N ratio spectra obtained with \textit{Suzaku}/XIS show a much higher complexity than the JEM-X and XRT spectra.
Two or three iron lines, an Fe edge and a strong soft excess are present in these spectra (see also \cite{miyazawa09}). As the main focus of our work is
the hard X-ray emission of \ngc, we restricted the fit to the data above 1.5~keV, in order to avoid the soft excess component,
we fixed the energy of the lines and edge 
($E_{\rm Fe I} = 6.4 \, \rm keV$, $E_{\rm Fe XXV} = 6.7 \, \rm keV$, $E_{\rm Fe XXVI} = 6.97 \, \rm keV$, $E_{\rm Fe \,\, edge} = 7.1 \, \rm keV$, 
in the source rest frame; Fig.~\ref{spe_ngc}) and we do not further discuss here the properties of these features.
Beside the {\sc pexrav} model, the \textit{Suzaku} data have been also fitted with a more physical Comptonisation model ({\sc compPS}).
In this case, we used a slab geometry 
and seed photons from a multicolor disc with a temperature fixed to $kT_{\rm bb} = 0.1 \, \rm keV$.

A cross-calibration factor has been added to all fitting models. The normalization of PIN data relative to XIS data was fixed to $C_{\rm PIN}=1.16$ (observations 
at the XIS nominal position\footnote{\emph{http://heasarc.nasa.gov/docs/suzaku/analysis/abc/node8.html}}). The cross-calibration factor of JEM-X relative to ISGRI
has been fitted for the strictly simultaneous, on-axis observation (July 2003), resulting in $C_{\rm JEM-X} = 0.9 \pm 0.2$, and then it was fixed to this value for 
all other observation periods.
The ISGRI cross-calibration factor relative to XRT was left free to vary and assumed the values 
$C_{\rm ISGRI} = 1.9{+0.9 \atop -0.6}, \,\, 0.7 \pm 0.3, \,\, 1.0{+0.4 \atop -0.3}$, for the May-June 2009, Jan 29--30, 2010, and the total XRT+ISGRI observations, 
respectively. \\
Errors quoted in this work are at the 90\% confidence level.

\subsection{Comptonisation features and spectral variability} 
\begin{table}[!t]
\hspace{-1cm}
\begin{minipage}[t]{\linewidth}
\begin{center}
\begin{tabular}{c c c c c c c c}
\hline\hline                
Instr.  & Obs. & model & $N_{\rm H}$               & $\Gamma$/$\tau$ & $E_{\rm C}$/$kT_{\rm e}$ & R & $\chi^2_{red} \rm \, /dof$  \\   
	     &        	  &	  & $[10^{22} \,\rm cm^{-2}]$ &                 & [keV]                    &   &                              \\   
\hline                      
\noalign{\smallskip}
J2+ISGRI     & July 2003        & pexrav & 3.5$\pm$2.5	 		  & 2.3 $\pm$ 0.2   	      & -  		      & 2.3${+2.7 \atop -1.4}$  & 0.9/11 \\
XIS+PIN	     & Aug 8--10, 2006  & pexrav & 3.28$\pm$0.07 		  & 2.01 $\pm$ 0.04 	      & - 		      & 1.0 $\pm$ 0.2 		& 1.04/271 \\
	     &   	        & compPS & 3.12$\pm$0.07 		  & 0.6 $\pm$ 0.2 	      & 81 $\pm$ 14 	      & 1.4${+0.3 \atop -0.2}$ 	& 1.09/270 \\
XIS+PIN	     & Aug 11--12, 2006 & pexrav & 3.35$\pm$0.07 		  & 2.03 $\pm$ 0.04 	      & - 		      & 0.9 $\pm$ 0.2		& 0.91/270 \\
	     & 		        & compPS & 3.20${+0.07 \atop -0.06}$      & 0.7 $\pm$ 0.2 	      & 70 $\pm$ 12 	      & 1.3 $\pm$ 0.2 		& 1.02/269 \\
XIS+PIN      & Jan 2007	        & pexrav & 3.17$\pm$0.07 		  & 1.93 $\pm$ 0.04 	      & - 		      & 0.6${+0.2 \atop -0.1}$ 	& 1.06/270 \\
             &  	        & compPS & 3.06$\pm$0.07		  & 1.0${+0.3 \atop -0.2}$    & $64 {+17 \atop -16}$  & 0.9 $\pm$ 0.2 		& 1.11/269 \\
J1+ISGRI     & Jan 2008	        & pexrav & 3		 		  & 1.8${+0.2 \atop -0.3}$    & - 		      & 1 			& 0.8/7 \\
J1+ISGRI     & June 2008        & pexrav & 3		 		  & 2.1 $\pm$ 0.2 	      & - 		      & 1 			& 0.61/13 \\
	     &        	        & pexrav & 3		 		  & 2.1 $\pm$ 0.2 	      & $> 40$ 		      & - 			& 0.65/12 \\
XRT+ISGRI    & May--June 2009   & pexrav & 2.6$\pm$0.2	  	          & 1.67${+0.1 \atop -0.09}$  & - 		      & 1 			& 0.84/202 \\
J2+ISGRI     & Jan 14--25, 2010 & pexrav & 3		 		  & 2.1${+0.1 \atop -0.2}$    & - 		      & 1 			& 0.5/8 \\
XRT+ISGRI    & Jan 29--30, 2010 & pexrav & 2.5${+0.4 \atop -0.3}$	  & 1.4 $\pm$ 0.2 	      & $36 {+85 \atop -15}$ & $<1.1$ 			& 1.16/79 \\
\noalign{\smallskip}
\hline                         
\end{tabular}
\end{center}
\hspace{0.85cm}
\begin{minipage}[t]{\linewidth}
\caption{Best fit parameters for the 9 data sets of \ngc\ analysed in this study.
	All fits include the additional component {\sc wabs} to account for absorption, and, when fitting the \textit{Suzaku} spectra, 
	3 Gaussian lines and an edge have been added to obtain the final model.
         When no uncertainties are reported, the parameter has been fixed to the corresponding value during the fitting procedure.
	 When no provided, the values of reflection and high-energy cut-off have been fixed to $R=0$ and $E_{\rm C} = 800 \rm \, keV$,
	 i.e. corresponding to no reflection and no cut-off in the fitted range.
	 }
\end{minipage}
\label{tab}
\end{minipage}
\end{table}
At hard X-rays, \ngc\ shows flux variations up to a factor of 2 on time scales of years (Fig.~\ref{lc_ngc}). 
In order to investigate the variability of the source parameters across the 7 years of observations, and to have comparable results in the different time periods, 
we fitted all the spectra with a {\sc pexrav} model modified by intrinsic absorption (fixed to $N_{\rm H} = 3 \times 10^{22} \rm \, cm^{-2}$ where not constrained).
During none of the nine observational periods it is possible to simultaneously constrain both the high-energy cut-off and the reflection fraction 
$R$ \footnote{$R$ is defined here as the relative amount of reflection compared to the directly viewed primary continuum}.
For all but one period (Jan 29--30, 2010), the best fit is achieved when assuming no high-energy cut-off and a reflection component either free to vary
($R = 0.6-2.3$) or fixed to the average value of $R=1$.
For the Jan 29--30, 2010 spectrum, a cut-off at $E_{\rm C} = 36 {+85 \atop -15} \, \rm keV$ and an upper limit at $R < 1.1$ for the reflection fraction provide the 
best fit.
For the June 2008 spectrum, a slightly worse fit, but still with $\chi^2_{\rm red} << 1$, is obtained with no reflection and a lower limit at 
$E_{\rm C} > 40 \, \rm keV$ for the high-energy turnover.
Only when fitting the total XRT$+$ISGRI spectra, the high-energy cut-off and the reflection fraction can be simultaneously measured, resulting in 
$E_{\rm C} = 113{+106 \atop -37} \, \rm keV$, $R = 1.2{+1.9 \atop -0.9}$. 
\begin{figure}[!b]
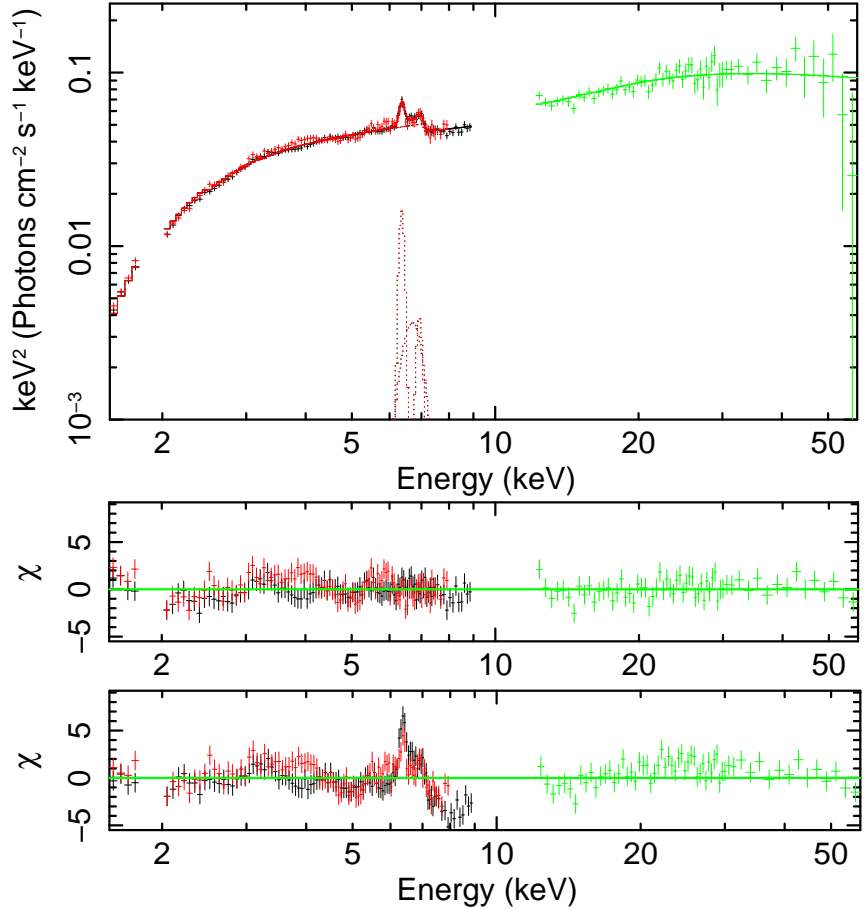

\begin{minipage}[t]{17cm}
 \hspace{1.2cm}
  \begin{minipage}[t]{7cm}
  \includegraphics[width=1.0\textwidth,angle=-90]{fig2.ps} 
  \end{minipage}
 \vfill
 \begin{minipage}[t]{7cm}
  \hspace{1.41cm}
  \psfig{width=.353\textwidth,angle=-90,figure=fig3.ps}
 \end{minipage}
 \vfill
 \begin{minipage}[t]{7cm}
  \hspace{1.41cm}
  \psfig{width=.434\textwidth,angle=-90,figure=fig4.ps}
 \end{minipage}
\end{minipage}
  \caption{\textit{Top}: \textit{Suzaku} XIS$+$PIN spectra of \ngc\ during the August 8--10, 2006 observation (XIS front-illuminated units in black, back-illuminated in red). 
  	   The spectra are fitted with a pexrav model, modified by intrinsic absorption, to which 3 gaussians at 6.4, 6.7 and 6.97 keV and an iron edge at 7.1~keV have been added. 
	 \textit{Center}: Residuals to the fit with the above model. 
	 \textit{Bottom}: Residuals to the fit with a simple absorbed power law model.
	   }
  \label{spe_ngc} 
\end{figure}
\begin{figure}[!t] 
\hspace{4.5cm}
\includegraphics[width=.4\textwidth,angle=0]{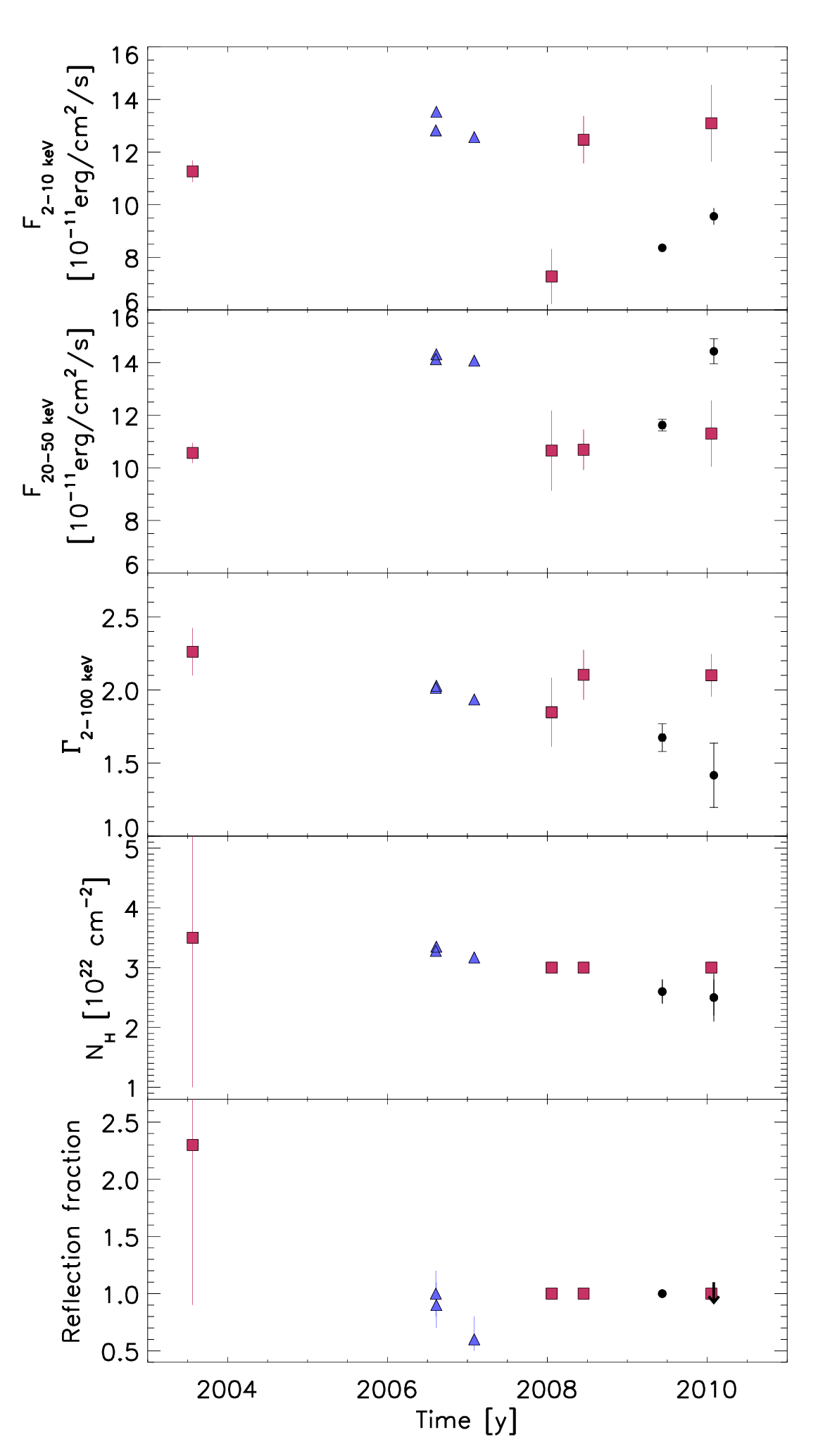} 
\caption{Evolution of the spectral parameters with time, as measured by \textit{Suzaku} (blue triangles), \textit{INTEGRAL} and \textit{Swift} (JEM-X$+$ISGRI red squares; 
         XRT$+$ISGRI black circles). The spectra have been fitted with a pexrav model with $N_{\rm H}$ fixed to $3 \times 10^{22} \, \rm cm^{-2}$ and R fixed to 1 when it is not possible 
         to constrain them. Fluxes are unabsorbed.
	 } 
\label{evolution} 
\end{figure} 

The intrinsic absorption shows limited variations in the range $N_{\rm H} = 2.5-3.5 \times 10^{22} \, \rm cm^{-2}$, whereas the power law photon index
varies between $\Gamma = 1.4$ and $\Gamma = 2.3$, with an average value of $\langle \Gamma \rangle = 1.9 \pm 0.3$ (Fig.~\ref{evolution}).
The only constraining values for the reflection fraction can be obtained with the \textit{Suzaku} spectra (Fig.~\ref{spe_ngc}), providing $R = 0.6-1$, when the spectra are 
fitted with the {\sc pexrav} model, and $R=0.9-1.4$ when the {\sc compPS} model is used.
With the latter model, we find values of the optical depth and temperature of the plasma cloud of $\tau = 0.6-1.0$ and 
$kT_{\rm e} = 64-81 \, \rm keV$, respectively, which translate into a high-energy cut-off at $E_{\rm C} \approx 120-160 \, \rm keV$ in the photon spectrum.
It is important to notice that the {\sc compPS} model does not provide a statistically better fit than the {\sc pexrav} model for the 3 \textit{Suzaku} spectra
and that the relatively low upper energy boundary of the PIN spectra ($\sim 50 \, \rm keV$) does not make this instrument the most suitable one for
measuring turnovers around 100~keV.
A more exhaustive analysis of the mean spectral properties of this source, exploring a wider parameter space and including \textit{XMM-Newton} data, 
will be presented in Lubinski et al. (in preparation).

\section{Discussion and conclusions}

The intrinsic hard X-ray spectrum of \ngc\ measured with \textit{INTEGRAL}, \textit{Swift} and \textit{Suzaku} data is well represented by a power law 
with $\Gamma = 1.9$, modified by a reflection fraction $R = 1$ and a high-energy cut-off at $E_{\rm C} \geq 40-110 \, \rm keV$.
During previous observations with \textit{BeppoSAX}, \textit{XMM-Newton} and \textit{INTEGRAL} \cite{matt01,bianchi03,dadina07,beckmann09},
this NLS1 showed similar properties to those revealed by our study, with a spectrum with photon index $\Gamma \simeq 2$, a strong Compton reflection with $R = 1-3$ 
but only lower limits at 100--200~keV for the high-energy cut-off. 
In spite of the observed flux variations, \ngc\ spectral shape seems to be quite stable, with only a hint of hardening of the spectrum and decreasing intrinsic
absorption during the recent hard X-ray observations. 

The intrinsic spectrum of \ngc\ is found to be consistent with that observed in average for NLS1.
An early work with a limited sample of 5 NLS1 detected with \textit{INTEGRAL} found an average steep spectrum at hard X-rays with $\Gamma \approx 2.6$, interpreted as due
to the presence of an unconstrained cut-off at $E_{\rm C} \leq 60 \rm \, keV$ \cite{malizia08}.
More recent results on a sample of 14 hard X-ray selected NLS1 have been presented by \cite{ricci11a,ricci11b,panessa11}. The average hard X-ray NLS1 spectrum 
results to be only slightly softer, with $\Gamma_{\rm NLS1} = 2.0-2.3$, compared to that of BLS1 ($\Gamma_{\rm BLS1} = 1.7-2.2$ \cite{ricci11a,panessa11,molina09}).
This is again possibly due to the presence of a cut-off in the NLS1 hard X-ray spectrum at lower energies of $E_{\rm C} \sim 50-60 \rm \, keV$ \cite{ricci11a,ricci11b}
compared to the values of $E_{\rm C} \geq 100 \rm \, keV$ observed in BLS1 \cite{ricci11a,ricci11b,molina09} and Seyfert 1 galaxies in general 
($E_{\rm C} \sim 90 \rm \, keV$; \cite{beckmann09}).

A quite large discrepancy is found in the estimate of the black hole mass of this object, with values ranging from
$2 \times 10^6$ to $10^8 \, \rm M_{\odot}$ \cite{hayashida98,papadakis04,bian07,nikolajuk09}. With a 0.1--300~keV luminosity of 
$L = 6 \times 10^{43} \, \rm erg \, s^{-1}$ and assuming a twice as large bolometric luminosity, \ngc\ presents an Eddington ratio of 
$L_{\rm bol}/L_{\rm Edd} = 0.01-0.5$. A black hole of few $10^6 \, \rm M_{\odot}$ fits better the scenario of \ngc\ belonging to a class of AGN in an early 
evolutionary state and powered by extreme accretion rates. In addition, a low mass would also agree with the lack of strong relativistic broadening of the broad 
Fe K$\alpha$ line recently detected in this AGN \cite{guainazzi10}.

\section*{Acknowledgments}
\small{ 
S.S. acknowledges the Centre National d'Etudes Spatiales (CNES) for financial support.
Part of the present work is based on observations with \textit{INTEGRAL}, an ESA project with
instruments and science data centre funded by ESA member states with the participation of Russia and the USA.
}

\end{document}